# Functional Unit: A New Perspective on Materials Science Research Paradigms

*Caichao Ye [1,2], Tao Feng [1], Weishu Liu [1,\*], Wenqing Zhang [1,\*]*

[1] *Department of Materials Science and Engineering & Institute of Innovative Materials, Southern University of Science and Technology, Shenzhen, 518055, PR China*

[2] *Academy for Advanced Interdisciplinary Studies & Guangdong Provincial Key Laboratory of Computational Science and Material Design, Southern University of Science and Technology, Shenzhen, 518055, PR China*

[\*]*Corresponding Authors: W.S. Liu:* liuws@sustech.edu.cn*; W.Q. Zhang:* zhangwq@sustech.edu.cn

**Abstract:**

New materials have long marked the civilization level, serving as an impetus for technological progress and societal transformation. The classic structure-property correlations were key of materials science and engineering. However, the knowledge of materials faces significant challenges in adapting to exclusively data-driven approaches for new material discovery. This perspective introduces the concepts of functional units (FUs) to fill the gap in understanding of material structure-property correlations and knowledge inheritance as the "composition-microstructure" paradigm transitions to a data-driven AI paradigm transitions. Firstly, we provide a bird's-eye view of the research paradigm evolution from early "process-structure-properties-performance" to contemporary data-driven AI new trend. Next, we highlight recent advancements in the characterization of functional units across diverse material systems, emphasizing their critical role in multiscale material design. Finally, we discuss the integration of functional units into the new AI-driven paradigm of materials science, addressing both opportunities and challenges in computational materials innovation.

# 1. Introduction

Materials probably shape our daily lives more than most of us realized, profoundly influencing transport, housing, clothing, communication, and food production. The new materials even marked the civilization level, including the Stone Age, Bronze Age, Iron Age, and Silicon Age (or Plastic Age), serving as an impetus for technological progress and societal transformation. The knowledge of materials science plays a fundamental role in supporting the discovery of new materials from the early activity of metallurgy, sintering of delicate china, and fabrication of cotton cloth. However, materials science rose as a discipline quite late until Northwestern University, Illinois, USA, founded the Department of Materials Science and Engineering in 1959 [1-2]. Materials science encloses partial knowledge of metallurgy, ceramic engineering, chemical engineering, and solid-state physics, and later shifts focus on the relationship between the microstructure and the physical/chemical properties [3]. The insight into the structure-properties relationship marked a great success in exploring new materials with high strength, high toughness, and excellent capability to serve in high temperatures. The new microstructures also realize the many counterintuitive new materials, e.g. nano-twinned cooper breaking the inversion connection between strength and toughness [4]. Recently, a dual-phase silicon nitride realized a tensile deformation, manifesting the importance of the microstructure [5]. The new materials process was critical in exploring the new microstructures for metals, ceramics, and polymers. The different cooling speeds were used to fabricate Fe-C alloy with different phases for different purposes. The sintering temperature is a dominant process parameter for the porosity of the ceramics. On the other hand, the degree of polymerization is also sensitive to the process parameters, such as temperature, time, solvent, etc. The scope of material science then developed and formed the discipline of "process-structure-properties-performance", which is also visualized by the tetrahedron to emphasize the more complicated connections. Microstructure control is a unique art of materials scientists. Recently, J.X. Huang added a frame for structure by the time-scale and length scale as we discuss the four components of materials science [6]. The structure's time and length scales further emphasize the process's importance of the process. The past years have greatly succeeded in various performance advances in the structure materials of metals, ceramics, polymers, and newly raised nanomaterials for many functions and applications through process-driven innovation. The integrated circuit chips' great success, along with Moore's Law, manifested the process-dominant innovation.

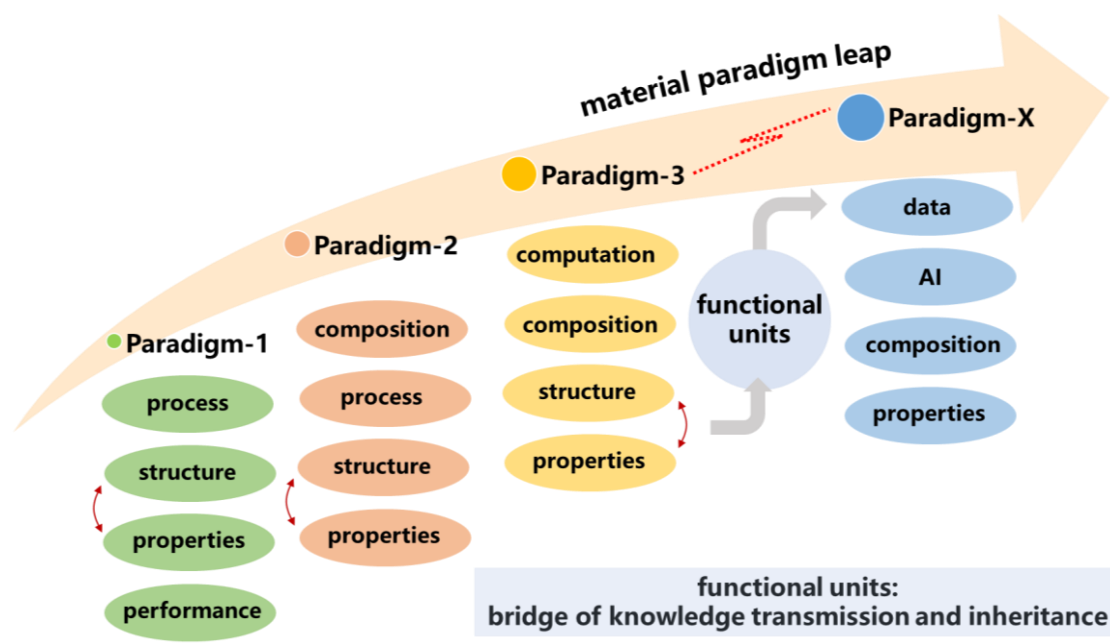

**Figure 1.** Materials research paradigm evolution and knowledge inheritance gaps due to a paradigm leap

As we expanded our scope from classic structural materials to functional materials, the significance of composition became increasingly evident. For instance, minor dopants such as phosphorus (P) or boron (B) determine the electric functions of silicon-based semiconductors, with doping content down to 10 ppm in a few nanometers thin layer from the surface, serving as the nano-size bricks for the microcircuits and changing our information communication route. Piezoelectric materials like PbZrO-PbTiO$_3$ often exhibit high mechanical-to-electrical responses near the morphotropic phase boundary, which is highly sensitive to the composition ratio between zirconium (Zr) and titanium (Ti). We can also find abundant examples of composition sensitivity in thermoelectric materials with high figure-of-merit, high-capacitance cathode materials, and high-activity single-atom catalysts, etc. Moreover, in organic materials, composition and its corresponding molecular connectivity sequence can be even more dominant in determining their physical and chemical properties than inorganic materials. The efficiency of organic photovoltaic materials (such as the P3HT:PCBM system) is directly related to their composition ratios and molecular arrangement [7]. With the tool of high-resolution TEM, people also widely identified the morphology of precipitates as highly sensitive to the minor alloying elements in the various metal alloys, emphasizing the importance of the composition in materials science.

On the other hand, quantum theory sets a new theoretical foundation for understanding the physical and chemical properties or experimentally measurable properties of microscopic atomic components in a given spatial arrangement and their interactions. Based on quantum theory, the first-principal computation initiated the interdisciplinary subfield spanning materials science and engineering, significantly enlarging the composition-driven materials innovation paradigm. We might summarize this material research paradigm as "computation-composition-properties-performance". Material process is streamlined into a tool for synthesizing materials with designed crystalline structure, rather than an art of controlling microstructures. From the 1990s, together with the significant progress of the computer, there was a growing interest in searching for new materials through trial-and-error methods or utilizing calculations based on first principles, density functional theory (DFT), molecular dynamics (MD), Monte Carlo techniques, and phase-field method[8-11]. These efforts culminated in the establishment of new materials databases [12], including the Materials Project [13], AFLOW [14], OQMD [15], and MatHub-3d [16]. Cubuk et al screened 12,831 lithium-containing candidate crystalline solids from the Materials Project database, with the goal of identifying those with high structural and chemical stability, low electronic conductivity, and low cost. Using a logistic regression-based ionic conductivity classification model, they identified candidate structures likely to exhibit fast lithium conduction, ultimately reducing the number of potential electrolyte materials from 12,831 to 21 that show promise as electrolytes, few of which have been examined experimentally [17]. Recently, Google DeepMind reported the use of materials databases to train Graph Networks for Materials Exploration (GNoME), which has predicted the structures of 2.2 million new materials, with only about 700 of them having been synthesized in laboratories [18]. We have established the Materials Hub with Three-Dimensional Structures (MatHub-3d) repository, MatHub-3d has driven significant advancements in thermoelectric materials, specifically in high-throughput (HTP) calculations of transport properties and material design. In some studies, we have investigated the relationship between transport properties and chemical bonds for particular types of thermoelectric compounds, using HTP results to enhance our fundamental understanding of these compounds [16, 19]. Noted that most of these new matters have not yet been synthesized. In other words, their distinguished properties and potential performance await future verification before we can call them materials.

The initial two decades of the 21st century have seen the emergence of big-data-driven science, which has also further shaped the paradigm of discovering new materials [20-21]. Estimates say there are $10^{108}$ potential organic molecules [22], which is far beyond what we have known from experiments and calculations. The data-driven materials innovation goes beyond quantum theoretical basis, alternatively use the information of the elements as descriptor for the target material properties, such as atomic size, ionization potential, electron affinity, electronegativity, etc. [23-24]. Scheffler et al. utilized big data and machine learning to construct a multi-dimensional descriptor feature space based on the properties of elements, such as ionization potential, electron affinity, and electronegativity, revealing physically interpretable descriptor-energy of materials relationships for predicting promising materials [25]. Zhang et al. utilized high-throughput computations and machine learning techniques to investigate the relationships between elemental information, chemical bond strengths, electronic structures, and the thermoelectric transport properties of the thermoelectric materials, theoretically predicted and experimentally validated a new class of high-performance diamond-like thermoelectric materials [9, 26]. By filling the 4$d$ vacancies in the interstitial sites of half-Heusler materials, they achieved the compositional interval between half-Heusler and full-Heusler, obtaining novel half-Heusler-like structural materials with unique physical properties [27]. Walsh et al. provided a detailed summary of the applications of machine learning in molecular and materials science, including the use of sufficient data and information on composition and elemental features to predict material properties [28]. The emergence of high-throughput computations and big-data-driven science has provided a vast data foundation for understanding the relationships between elemental properties, crystal structures, and material properties. While researchers have traditionally used the "composition-crystalline structure-properties-performance" paradigm to enhance material performance by refining elemental composition and microstructure [6], the limitations of this approach are becoming more apparent as material systems grow in complexity. The increasing number of atomic-scale features leads to an exponential rise in variables, making it difficult to establish effective correlations at both macroscopic and microscopic levels. Additionally, this paradigm relies heavily on classical theories focused on altering compositions and microstructures to optimize performance. However, a significant gap remains in understanding how these changes correlate with macroscopic performance, especially in terms of new property formation mechanisms. This gap not only challenges the prediction and design of new material performance but also limits the exploration of novel properties and physical effects (**Figure 1**).

The structure-property relationship is integral to the disciplinary essence of materials science. However, with the rapid paradigm shifts in materials research, the big-data and AI-driven paradigm has not adequately preserved the materials knowledge of the "relationship between structure and properties". Here, we refer to it as the "Inheritance Rift of Materials Science." In this perspective, we aim to illuminate the bridge between the previous paradigms and the new paradigm driven by big data and AI, with a focus on the new concept of "Functional Units (FUs)" that defines a microscopic relationship between structure and properties.

## 2. The Concept Evolution of Functional Units

Atoms are submicroscopic particles that construct our matter world and enrich our life. Crystalline structures for inorganic materials and covalently linked chains for chemical formulas for organic materials are fundamental for the relationship between structure and properties in the field of materials science and engineering. The past years have witnessed a vast number of new materials that were synthesized or predicated. Researchers also noted that some physical properties are more sensitive to a group of atoms, or atomic clusters, or single atoms in special chemical surrounding. One of the early examples could be C. T. Chen's anionic group theory for the nonlinear optical susceptibility of the crystals in the 1980s [29]. Their following works predicated the $(B_3O_6)^{3-}$ planar group as a critical component group for new nonlinear optical materials [30-31]. Here, we refer to it as a type of "functional unit", which means the unit of $(B_3O_6)^{3-}$ has a function of nonlinear optical properties. Another functional unit could be nanotwinned units for exceptional mechanical properties. In 2009, Lu et al. reported nanoscale twined boundaries in pure Cu realized significant strength enhancements without loss of ductility.[32] The concept of highly strong and ductile nanotwinned units were later verified to be effective in ceramic materials or carbon materials. Tian et al employed nanotwinned units and appropriate hierarchical structuring ideas, successfully synthesizing extremely hard nanotwinned cubic boron nitride (cBN) in 2013 [33], and designing and synthesizing nanotwinned diamond in 2014 [34]. The nanotwinned diamond had twice the hardness of single-crystal diamond, setting a new record in ultra-hard materials. In functional materials, doping or alloying are well-known strategies for regulating physical or chemical properties and achieving exclusive functions. The journey to search for more effective "rattling dopants", for the filled skutterudites or clathrate for low thermal conductivity while high electrical conductivity, could be another example to interpret the idea of the functional unit. Nobly, more examples could be outlined [35], we would like to emphasize that the idea of functional units serves as a new insight for the materials scientist in China.

Since 2009, China governments have successively launched several research projects, including the major research project "Structure Design and Controllable Preparation of Function-Oriented Crystalline Materials" (NSFC, 2016), and the "Fundamental Research on High-Performance Materials with Functional Unit and Structural Architectures" (NSFC, 2019). These projects have significantly stimulated the research to reconsider the research paradigm of materials science in China, significantly advancing nonlinear optical (NLO) materials [36-40], magnetic organic materials [41-43], and thermoelectric materials [44-46]. Furthermore, some databases relative to the functional units are built up. For example, Guo et al. have developed a functional unit database for nonlinear optical materials and catalytic materials (www.functionalmotif.com), and Du et al. used a topological scaling algorithm to identify charged two-dimensional units in the Materials Project database and constructed a 2D ionic functional unit database (www.n11.iphy.ac.cn/databases) [47]. Recently, our team has constructed $Ag_2Te_{1-x}S_x$-based materials in which amorphous functional unit coexist with crystalline functional unit. By leveraging the low thermal conductivity and high flexibility of the amorphous blocks, together with the high electrical performance of the crystalline blocks, we have successfully obtained semiconductor glass with superior flexibility and high room temperature thermoelectric performance [48]. Besides, we designed Ru-based 4c/4d disordered sublattice functional unit and obtained strange glass-like thermal transport behavior with unusually low lattice thermal conductivity (~1.65 $Wm^{-1}K^{-1}$ at 340 K) in $TiRu_{1.8}Sb$, being the lowest among reported half-Heusler thermoelectric materials [49].

We can conclude that the "functional unit" refers to the microscopic atomic unit that plays a crucial role in the macroscopic functionality of materials. The functional unit is an intermediate material unit with specific functions introduced between the atomic/molecular scale and the macroscopic scale. It can encompass specific atomic configurations, molecular interactions, and localized structures on surfaces and interfaces. Unlike the crystalline cell unit in metals, ceramics, and semiconductors, or the monomer in organic polymers, the functional unit simplifies the complex relationships between structure and properties, aligning with the Chinese philosophical principle of "the greatest truth comes simplest". It simplifies the relationships between structure and properties and provides a bridge to the new paradigm driven by big data. functional units bridge the gap between the classic materials research paradigm and the new trend driven by big data, enabling rapid advancements in material science by focusing on the configuration of functional units, structural architectures, and their interactions.

## 3. High-Performance Materials with Functional Unit and Architecture

In the past years, I have witnessed great interest in identifying functional units across various materials. Two excellent review articles have recently provided comprehensive summaries of this field [50-51]. Here, we aim to discuss functional units from a new perspective, focusing on their size scale and manipulation strategies.

**3.1 Scale of Functional Units**

The scale of functional units refers to the dimensions or ranges at which these units operate in materials science and engineering, spanning from the microscopic, mesoscopic to the macroscopic scale. Their functions encompass various aspects such as magnetic, electrical, thermal, mechanical, and optical properties, as summarized in **Figure 2**.

At the microscopic scale, functional units often start at the atomic or molecular level, where specific lone-pair electrons, atomic configurations, molecular properties, and chemical bonding play a crucial role. For example, the atoms with lone pair electrons, serve as the smallest functional units, impelling the phonon anharmonicity, which is benefic for ultralow lattice thermal conductivity [13, 52-53]; single-molecule/ionic magnetic units possess extremely high magnetic properties [42-43]; linear three-atoms low thermal conductivity resonant units can significantly reduce heat conduction [45-46], and nonlinear optical anionic units exhibit outstanding nonlinear optical performance [54-55].

At the same time, inorganic sub-lattice units that couple thermal, electrical, and magnetic functions are also noteworthy [27].

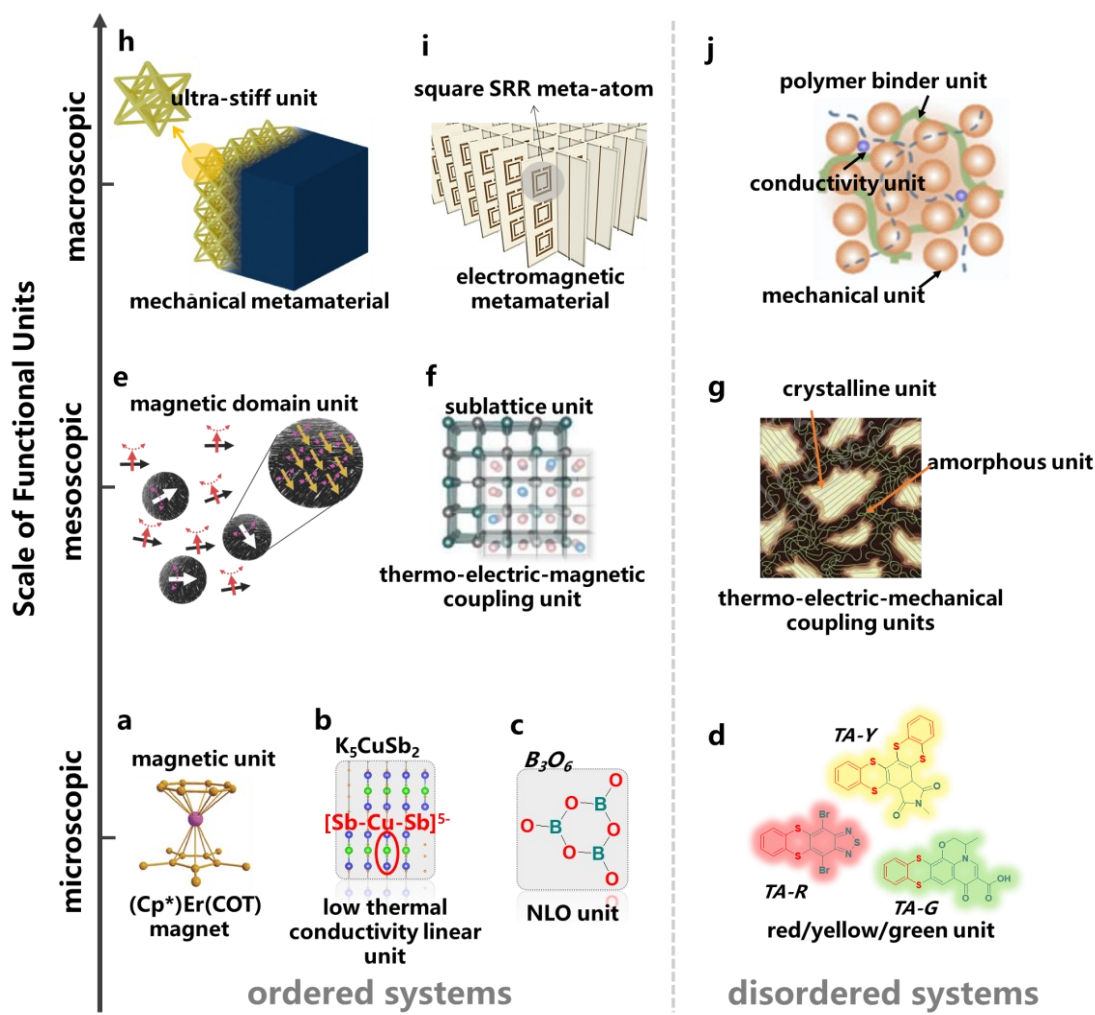

**Figure 2.** Scale and properties of functional units. Reprinted (adapted) with permission from Ref.41, Copyright 2011 American Chemical Society; Ref.59, Copyright 2022 American Chemical Society

At the mesoscopic scale, nano-precipitates or nano particles in polycrystalline materials or microdomains in polymers could be the functional units[56], which are crucial for modulating a wide array of material properties, including mechanical, magnetic, thermoelectric, etc., as well as enhancing the stability of materials. For instance, the intrinsic nano-precipitates or externally mixed nanoparticles also significantly influence the strength and ductility of polycrystalline materials, while the microdomain structures in polymers can control phase separation and self-assembly behavior. There are widely experimental results that verify the intrinsic nano-precipitates or externally mixed nanoparticles could decouple the transport of phonons and electrons, increasing the thermoelectric figure of merit. In 2017, Zhao et al. reported the single-domain soft and hard magnetic Co nanoparticles as functional unit in the thermoelectric material of $Ba_{0.3}In_{0.3}Co_4Sb_{12}$, showing an exceptional charge transfer from the magnetic inclusions to the matrix and multiple scattering of electrons by superparamagnetic fluctuations, in addition to phonon scattering. [57-58]. Furthermore, the distribution of functional units at the mesoscopic scale can significantly affect the macroscopic performance of materials. For example, by controlling the distribution of nanoparticles, composite materials can achieve excellent thermal and electrical conductivity [59]. The introduction of $Pt_3Ni$ or hydroxy-$BaCaSiO_4$ nanoparticle units can accelerate the proton desorption of electrocatalytic reaction intermediates, thereby improving the OER performance of electrocatalytic materials [60-61].

**Table 1** Summarized the partial functional units with an emphasis on their size scale.

| Functions | [Unit]-[Corresponding function] | Scale | References |
|---|---|---|---|
| Optical units | [$(B_3O_6)^{3-}$; In/$SiP_4$; $GaS_4$, $(GeO_3)^{4-}$]-[nonlinear optical] | microscopic | Ref.29, 37-38, 40 |
| | [-C=C- $sp^2$ structure]-[phosphorescence unit] | microscopic | Ref.62 |
| | [*TA-R/TA-Y/TA-G* molecules]- [luminescent units for red/yellow/green] | microscopic | Ref.63 |
| | [square SRR meta-atom]-[broad-band and low-loss for microwave] | macroscopic | Ref.64 |
| Thermal units | [Sb-Cu-Sb]-[low thermal conductivity] | microscopic | Ref.46 |
| | [Ru-based 4c/4d disordered sublattice]- [low lattice thermal conductivity] | microscopic | Ref.49 |
| | [$Bi_2Te_3$ layers]-[ low thermal conductivity] | mesoscopic | Ref.65-66 |
| | [$R_yCo_4Sb_{12}$]-[low thermal conductivity] | microscopic | Ref.67 |
| | [Cu-Se sublattice]- [low thermal conductivity] | microscopic | Ref.68 |
| Electrical units | [PEDOT:DAE interface polaron state]-[high mobility] | microscopic | Ref.69 |
| | [$EuTiO_3$]-[magnetodielectric coupling multiferroism] | microscopic | Ref.70 |
| | [Polymer-Unit]-[high carrier mobility] | microscopic | Ref.71-72 |
| | [carbon nanotubes]-[ conductive filler units for composite materials] | mesoscopic | Ref.59 |
| | [2D-nanosheet]-[high-rate ionic transport] | mesoscopic | Ref.73-75 |

| Mechanical units | [nanotwinned-Cu]-[high strength] | microscopic | Ref.4 |
| --- | --- | --- | --- |
| | [nanotwinned-Carbon]-[extremely hard] | microscopic | Ref.34 |
| | [nanotwinned-cBN]-[extremely hard] | microscopic | Ref.33 |
| | [silicon-based particles]-[mechanical units for composite materials] | mesoscopic | Ref.59 |
| | [octet microlattice]-[ultra-light and ultra-stiff] | macroscopic | Ref.76 |
| Magnetic units | [(Cp*)Er(COT); [(Cp$^{iPr5}$)Dy(Cp*)]$^+$]-[single-molecule/ion magnets] | microscopic | Ref.41-42, 77 |
| | [LSMO:NiO]-[low-field magnetoresistance] | mesoscopic | Ref.78 |
| | [MnTe layers]-[antiferromagnetic coupling] | mesoscopic | Ref.65-66 |
| | [FePt]-[magnetic anisotropy and magnetostriction] | mesoscopic | Ref.79 |
| | [Fe, Co, Ni particles]-[soft-magnet metal] | mesoscopic | Ref.57 |
| | [BaFe$_{12}$O$_{19}$ particle]-[ferromagnetism] | mesoscopic | Ref.58 |

In scaling functional units from the mesoscopic to the macroscopic level, the integration and development of functional units endow materials with unprecedented high performance. Various metamaterials have been developed through precise microstructural design, such as mechanical metamaterials and microwave metamaterials [80-81]. These metamaterials exhibit exceptional performance in specific application scenarios. For example, mechanical metamaterials can achieve lightweight and high-strength characteristics through the design of units with specific microstructures [76], and electromagnetic metamaterials on the other hand, go far beyond traditional materials, their remarkable properties are primarily determined by the shape, size, geometry, and precise arrangement of functional units that constitute them [80, 82], exhibit excellent performance in electromagnetic wave control, making them suitable for applications in stealth materials and microwave absorbers [64, 82]. Notably, the multiscale design of functional units involves the integration of properties and functions across these different scales. This approach allows for the regulation of material performance at both microscopic and macroscopic levels, thus enabling the development of high-performance materials with breakthrough and transformative properties.

## 3.2 Architecture Engineering of Functional Units

Architecture engineering of functional units redefines the classical "composition-process-structure-properties" framework of materials science research by introducing spatial hierarchy and directed organization as primary performance determinants [50]. This paradigm shift creates new frontiers in materials processing, demanding precise control over functional unit arrangements across scales. Below, we systematically demonstrate how architectural design principles govern material functionalities by using few selected examples.

In ordered systems at the microscopic scale, recent advances in thermoelectric materials highlight the importance of sublattice functional units and tailored architectures in optimizing thermal-electrical-magnetic coupling and thermal-electrical-mechanical coupling. For example, stepwise cooling of MnTe(s)/$Bi_2Te_3$(l) melts, annealing at the narrow temperature window above the crystallization point of $Bi_2Te_3$, and rapid water quenching, alternating MnTe and $Bi_2Te_3$ sublattice layers are induced to form ordered n-type $(MnBi_2Te_4)(Bi_2Te_3)_n$ 2D heterostructures (n = 1, 2, 3) [83]. Architectural precision in this lamellar system spatially decouples magnetic planes while establishing gradient-controlled magneto-thermoelectric synergy [66]. Increasing $Bi_2Te_3$ interlayer thickness (e.g., $MnBi_4Te_7$, $MnBi_6Te_{10}$) reduces antiferromagnetic exchange interactions ($J \approx 0.1$ meV for n=3 vs. $J \approx 0.5$ meV for n=1), tuning magnetic anisotropy without compromising carrier mobility. Quantum transport via narrow bandgaps and Dirac surface states further demonstrates structural engineering's impact [65]. Our team recently developed tunable 3D magnetic sublattice architectures via arc melting, mechanical alloying, and spark plasma sintering. By adjusting sintering temperatures, Fe/Cu diffusion is controlled to embed 4c/4d-site magnetic sublattices within TiSb frameworks, creating coherent 3D architectures with atomic-scale registry [84]. This processing creates a coherent 3D architecture characterized by Fe/Cu magnetic sublattice functional units incorporated in a TiSb sublattice functional units with atomic-scale registry. This method tailors thermo-electric-magnetic coupling by generating localized electronic states near the Fermi level that mediate interactions between magnetic moments and conduction electrons, inducing strong electronic correlations and novel magneto-thermoelectric functionalities. For thermal-electrical-mechanical coupling in Se-based compounds, Wei et al. reported the functional unit of dissipative chemical bond plasticity, which characterized by "easy gliding, high cleavage resistance, and low stiffness" [85]. Construct these plasticity functional units within high-performance brittle thermoelectric matrices by Bridgeman-Stockbarge method, achieving synergistic optimization of plasticity and thermoelectric performance through microscopic structural engineering [86]. In brittle $Ag_2Se$ matrices, when the plasticity functional units content reached 30–40%, materials exhibited both excellent plasticity and electrical transport properties, surpassing the inverted relationship predicted by traditional linear mixing rules [87]. This phenomenon is attributed to the formation of a connected network structure by the plasticity functional units. A series of $(Ag,Cu)_2(S,Se,Te)$-based thermoelectric materials achieved a room-temperature bending strain of 20% and a zT value of 0.6 [88]. The fabricated flexible thermoelectric devices demonstrated a maximum normalized power density of 30 $\mu W \cdot cm^{-2} \cdot K^{-2}$, approximately four times higher than that of conventional $Bi_2Te_3$-based rigid devices [89].

At the mesoscopic scale, magnetic functional unit distribution is precisely controlled through processing techniques. For instance, ordered Fe-based magnetic functional units can be incorporated into $Bi_2Te_3$-based thin films using screen-printing and hot-pressing methods. In this process, the screen-printing technique allows for the precise deposition of Fe-based magnetic functional units onto the $Bi_2Te_3$ film, while hot-pressing ensures the consolidation and adhesion of these functional units within the film matrix. This combination of techniques induces a specific ordered arrangement of the Fe-based functional units within the $Bi_2Te_3$ matrix, leading to a coherent 3D architecture [90-91]. The resulting structure exhibits enhanced electrical conductivity compared to disordered distributions, demonstrating the potential for optimizing material properties through architecture engineering. Similarly, in the combined method of DC/radio frequency magnetron sputtering, the FePt magnetic functional units are precisely incorporated into $Sb_2Te_3$ thin films, inducing robust magnetoelectric coupling effects. The DC magnetron sputtering technique allows for the precise deposition of FePt onto the $Sb_2Te_3$ film, while the radio frequency magnetron sputtering ensures the uniform distribution and adhesion of these units within the film matrix. The FePt units, located at specific sites, modulate the carrier transport properties, resulting in enhanced thermoelectric performance. This architectural modulation enables precise control over thermo-electric-magnetic coupling, optimizing the thermoelectric properties of the material. The localized electronic states near the Fermi level mediate interactions between local magnetic moments and conduction electrons, inducing strong electronic correlations and unlocking novel magneto-thermoelectric functionalities [79]. In the research of high-performance organic thermoelectric materials, the thermoelectric performance of materials can be effectively improved by regulating the ratio and distribution of polymer crystalline functional units and amorphous functional units. Pipe et al. pioneered a polar solvent treatment strategy to engineer phase segregation in PEDOT, creating hierarchical microstructures with optimized crystallinity. The process involves spin-coating PEDOT solutions in DMSO onto substrates, followed by immersion in methanol baths to induce controlled aggregation. This method produces a bicontinuous network of crystalline PEDOT domains embedded within an amorphous matrix. The ordered crystalline functional units enhance charge carrier mobility through π-π stacking, while the amorphous functional units suppress lattice thermal conductivity, achieving a record room-temperature zT of 0.42 [92]. Building on this, in 2015, Müller-Buschbaum et al. introduced ethylene glycol post-treatment to further align PEDOT chains into edge-on configurations, increasing the proportion of ordered lamellae. This structural refinement boosts electrical conductivity by facilitating intermolecular charge hopping while maintaining low thermal conductivity [93]. These advancements demonstrate that mesoscale control over crystalline-amorphous functional units enables simultaneous optimization of thermoelectric performance and mechanical flexibility.

Furthermore, at macroscopic scale, through the projection micro stereolithography technology, by curing photosensitive resin layer by layer to form a polymer template, and then performing subsequent processes such as degreasing and sintering, it enables the fabrication of octet micro-lattices from polymers, metals, and ceramics, producing exceptionally stiff, strong, and lightweight materials [76]. In microwave and terahertz metamaterials, spin-coating PMMA glue on a $SiO_2$ substrate and then using electron beam direct writing lithography technology to fabricate split-ring resonator patterns. The high-precision lithography-electron beam etching hybrid processes realize ordered split-ring resonator (SRR) array architectures, facilitating precise control over electromagnetic wave properties [82].

In disordered systems, configurational entropy plays a critical role. For instance, carbon-based quantum dots demonstrate a linear relationship between entropy and phosphorescence lifetime, enabling entropy-mediated control over excited-state dynamics [62]. Interfacial engineering in organic thermoelectric materials generates light-responsive "interface polaron state" units, highlighting the importance of interfacial entropy in optimizing performance [69]. Entropy, as a descriptor of architecture, influences microstructure, electronic states, and macroscopic properties [94-95]. Particularly in multicomponent systems, unit interactions coupled with entropy enable unprecedented structural optimization and property enhancement.

The representative cases substantiate that functional-unit architecture engineering serves as an effective strategy for modulating macroscopic properties (optical, electrical, magnetic, and multifunctional couplings) in advanced materials. Researchers can achieve enhanced material performance through precisely engineered inter-unit interactions or sub-lattice couplings and unlock new potential for advanced technological applications.

## 4. Limitations of the AI-derived Paradigm based on Elemental Information

Recent advancements in computational simulation methods and artificial intelligence have revolutionized materials science by enabling high-throughput exploration of structure-property relationships, significantly expanding the search space for advanced functional materials [18, 96]. The Materials Genome Initiative (MGI) launched in the early 21st century marked a paradigm-shifting transformation in materials research. G. Ceder and his collaborators established a systematic framework integrating high-throughput computation with data-driven methodologies for materials discovery. Its core objective focused on constructing a "computation-data-experiment" integrated system to establish quantitative structure-property relationships between material composition/structure and performance. This initiative propelled the global proliferation of materials databases, driving rapid advancements in high-throughput computational studies of inorganic crystalline materials, including electronic structure calculations, thermodynamic property predictions, and carrier transport simulations[97]. In the mid-2010s, materials informatics research was predominantly focused on specific crystal structure types, such as perovskites and zeolites, due to the inherent challenges in developing universal structural representations compatible with machine learning frameworks [28]. Researchers could leverage element-composition information and material property descriptors as built-in representations within the model by concentrating on a single structure type. The success of these approaches largely depended on the validity and relevance of the descriptors used. An adequate descriptor must be more straightforward to obtain than the target property and ideally be of as low dimensionality as possible [25]. However, with the rapid evolution of the AI paradigm in materials science, it has become increasingly evident that these traditional descriptors lack detailed structural information, such as atomic bonding and symmetry. This limitation poses significant challenges in building machine learning models capable of establishing genuine structure-activity relationships or enabling inverse materials design, going beyond mere property prediction for inorganic crystals and organic molecules [98-99], as shown in **Figure 3b**. Consequently, developing accurate structural representations and building robust structure-property relationship models are critical for understanding functional regulation mechanisms and designing advanced materials.

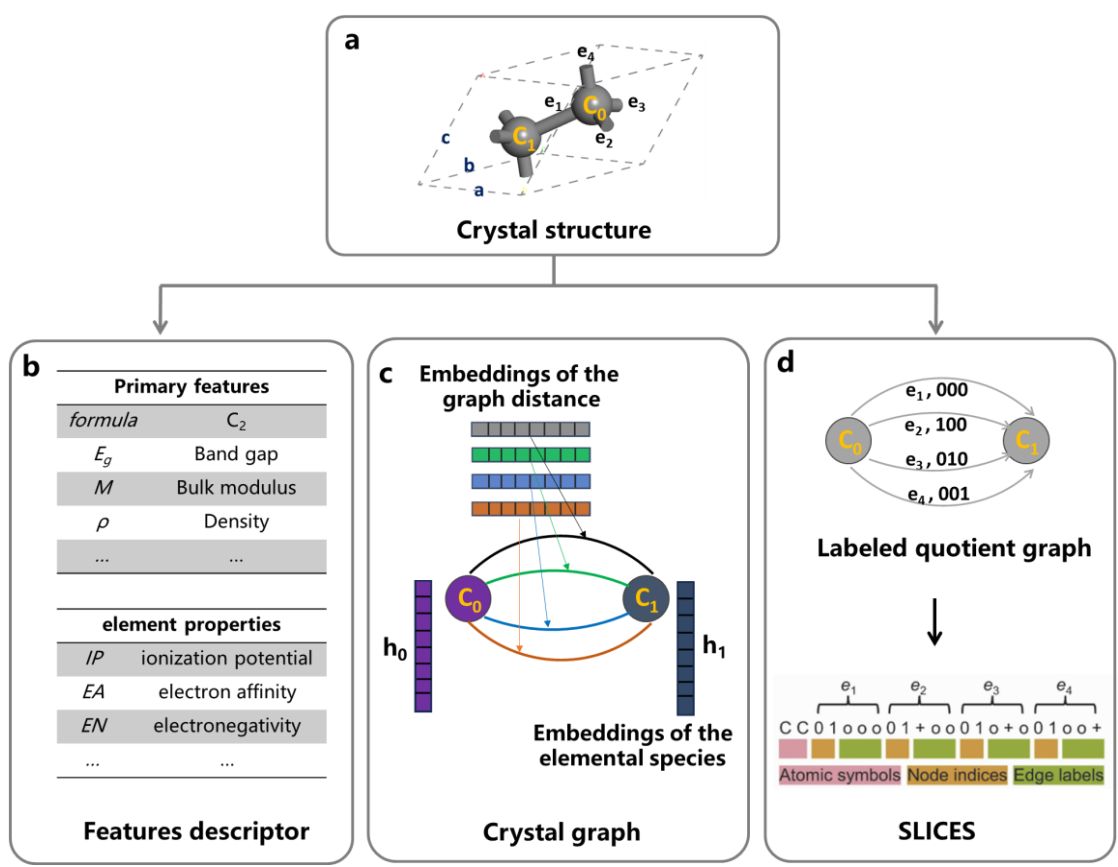

**Figure 3.** (a). Crystal structure; (b) Element-composition information and material property features descriptors; (c) Crystal graph representation for materials; (d) The labeled quotient graph serves as an intermediary to translate between crystal structures and SLICES strings.

Current structural representation approaches employ diverse encoding strategies to bridge molecular/crystalline systems with machine-interpretable formats. Widely adopted frameworks including molecular fingerprints, SMILES strings, and graph-based representations (molecular/crystal graphs) have demonstrated effectiveness in establishing structure-property correlations across organic molecules and inorganic crystals[18, 100-103], as shown in **Figure 3c**. Notably, structural fingerprints encode atomic configurations (e.g., bond lengths, coordination geometries) into machine-interpretable formats, enabling precise modeling. A notable innovation in crystalline materials is the SLICES framework (Simplified Linear Input Crystal Encoding System) developed by Xiao et al. [104], which translates crystal structures into topological graphs and compact text strings, as shown in **Figure 3d,** preserving both chemical composition and bond topology. While SLICES represents a breakthrough in crystalline material informatics, it remains constrained by the traditional "computation-composition-properties-performance" paradigm.

The necessity of knowledge inheritance in structural representation models for materials science arises from fundamental challenges in contemporary machine learning frameworks. Current structural representation models, despite their remarkable predictive accuracy for organic molecular systems and inorganic crystalline materials, face critical limitations knowledge extraction and inheritance: The implicit encoding of material knowledge in current structural representation ML frameworks (such as atomic local environment descriptors, property-weighted atomic features, and neural network convolutions) hinders both the inheritance of established material knowledge and the extraction of new scientific principles. Additionally, current structural representations prove inadequate for disordered material systems, including organic polymers and organic-inorganic composites, where structural complexity defies conventional encoding schemes.

Therefore, current research efforts are focused on developing more interpretable representation frameworks that align with materials scientists' cognitive patterns while maintaining transparent information encoding. A promising strategy involves mapping global material properties onto local structural entities (atom clusters, functional groups) to establish physically meaningful structure-property relationships. This limitation underscores the necessity for next-generation representations that systematically incorporate functional units as fundamental design variables - a transformative step toward design of novel materials.

## 5. Knowledge Inheritance of Materials Science by Functional Units

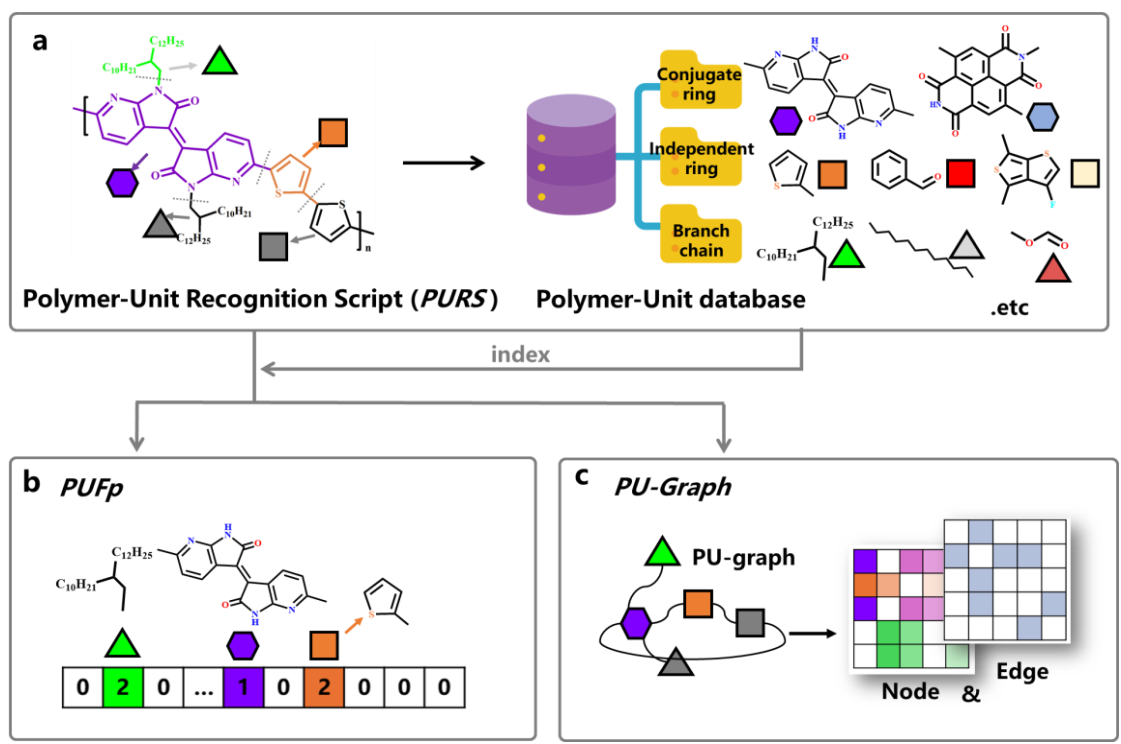

**Figure 4.** (a) *PURS* for identification of "Polymer-Units" from organic polymer functional materials, and Polymer-Unit database; (b) Polymer-Unit Fingerprint (*PUFp*) and (c) Polymer-Unit Graph (*PU-Graph*), accessible expression of polymer organic functional materials for machine learning.

The AI paradigm in materials science is undergoing a shift from traditional structure-property mapping to establish comprehensive relationships between functional units, structural architectures, and material properties [105]. This evolution aims to enhance model interpretability and knowledge inheritance, enable extraction of novel scientific principles, and facilitate the design of high-performance complex materials. Modern machine learning algorithms enable AI to identify promising combinations or architectures of functional units from vast experimental and computational data, accelerating the design of new materials. Additionally, AI models can predict the correlation between functional units and macroscopic material properties, such as magnetic, electrical, thermal properties, etc., and optimize material units through algorithmic approaches. The identification and characterization of key functional units have emerged as critical challenges in this new paradigm. Consequently, the development of FU-based ML frameworks, incorporating advanced structural representations, property-weighted unit features, and neural network models that include the concept of units, represents a strategic direction for AI-driven materials research in the foreseeable future.

Organic functional materials, especially polymers, are common high-performance complex materials composed of a series of precursor molecular "Polymer-Unit" through chemical reactions. Therefore, it is logical to consider "Polymer-Units" as the functional units of organic polymer materials. Recently, the authors have made significant progress in the development of FU-based ML frameworks. We proposed the concept of "Polymer-Unit" based on the structural characteristics of organic polymer functional materials and developed the *PURS* (Python-based Polymer-Unit Recognition Script) for automated identification of "Polymer-Units" from SMILES codes according to specified rules. Then a comprehensive "theoretical-experimental" Polymer-Unit database were established, encompassing organic semiconductor, photovoltaic, and dielectric materials, containing ≥400 characterized polymer units (**Figure 4a**) [72]. Additionally, the authors have also developed "Polymer-Unit Fingerprint" and "Polymer-Unit Graph" as two advanced structural representation methods specifically for polymer materials (**Figure 4b and 4c**). These representations fundamentally differ from conventional molecular fingerprints or graphs by explicitly encoding functional unit information, each bit or node in the "Polymer-Unit Fingerprint" or "Polymer-Unit Graph" represents a specific functional unit from the "polymer unit database" and edges representing architectural connections between units [106]. The framework enables concise yet comprehensive embedding of critical information (including unit structures, physicochemical properties, and inter-unit connections) within the node and edge matrices of the Polymer-Unit Graph, the polymer units enable significant and effective dimensionality reduction of molecular graph information [107], achieving a significant reduction in graph structural entropy while enhancing the interpretability of the network model. This ensures both completeness and minimalism in representing organic polymer architectures, thereby facilitating systematic analysis of the coupling relationships between polymer units, graph structural entropy, and macroscopic material properties. Furthermore, based on the identification of these "Polymer-Unit", the author developed a visual graph neural network model, *PU-LRP*, which projects model contribution weights for material properties (e.g., mobility, power conversion efficiency) onto specific Polymer-Units, enabling visual quantification of functional unit contributions to target properties, facilitating interpretable and visual analysis of machine learning models, and revealing the "functional units-structural architectures-property" relationships, thereby enable knowledge inheritance, extraction of novel scientific principles, and guiding the design and development of new organic polymer functional materials [71].

The materials research paradigm has shifted from traditional "elements-composition-crystal structures" analysis to a FU-based approach that directly maps to macroscopic properties. This transformation spans multiple scale, from molecular level to the microscopic and mesoscale, bridging critical gaps in understanding structure-property relationships as the field transitions from "composition-microstructure" paradigm to a big data-AI driven paradigm. In big-data & AI driven materials research paradigm, current research mainly emphasizes the developing mathematical descriptions of structural representations and physics-constrained neural network based on functional units. Researchers employ artificial intelligence to construct interpretable models that link functional units, structural architectures, and macroscopic properties, thereby enabling the knowledge inheritance of materials science through functional units.

## 6. Conclusion and Prospects

Data-driven AI paradigms are reshaping materials science and engineering. However, the black-box nature of AI models raises a new concern of knowledge inheritance of materials science. Functional units could be a promising choice to bridge the gap in interpreting structure-property correlations and maintaining knowledge inheritance as the "composition-microstructure" paradigm transitions to a data-driven AI paradigm. Addressing the engineering challenge to tune the ordering structures of functional units could impetus the progress of materials science and engineering.

**Notes**

The authors declare no competing financial interest.

**Acknowledgments**

Financial support was provided by the National Natural Science Foundation of China (No. 92463310, No. 92163212, and No. 52473235), Natural Science Foundation of China for Distinguished Young Scholars (No. T2425012), National Key R&D Program